\documentclass[twocolumn,prl,a4paper,superscriptaddress,showpacs,preprintnumbers,tightenlines]{revtex4}
\usepackage{amsmath}
\usepackage{graphicx}
\usepackage{epsfig}

\begin{document}

\preprint{\tighten\vbox{\hbox{\hfil Belle Prerpint 2003-6}}}
\preprint{\tighten\vbox{\hbox{\hfil KEK   Preprint 2003-19}}}

\title{Observation of {\mbox {$\boldmath{B^+\rightarrow
	\rho^+\rho^0}$} }} 
\affiliation{Aomori University, Aomori}
\affiliation{Budker Institute of Nuclear Physics, Novosibirsk}
\affiliation{Chiba University, Chiba}
\affiliation{Chuo University, Tokyo}
\affiliation{University of Cincinnati, Cincinnati, Ohio 45221}
\affiliation{Gyeongsang National University, Chinju}
\affiliation{University of Hawaii, Honolulu, Hawaii 96822}
\affiliation{High Energy Accelerator Research Organization (KEK), Tsukuba}
\affiliation{Hiroshima Institute of Technology, Hiroshima}
\affiliation{Institute of High Energy Physics, Chinese Academy of Sciences, Beijing}
\affiliation{Institute of High Energy Physics, Vienna}
\affiliation{Institute for Theoretical and Experimental Physics, Moscow}
\affiliation{J. Stefan Institute, Ljubljana}
\affiliation{Kanagawa University, Yokohama}
\affiliation{Korea University, Seoul}
\affiliation{Kyoto University, Kyoto}
\affiliation{Kyungpook National University, Taegu}
\affiliation{Institut de Physique des Hautes \'Energies, Universit\'e de Lausanne, Lausanne}
\affiliation{University of Ljubljana, Ljubljana}
\affiliation{University of Maribor, Maribor}
\affiliation{University of Melbourne, Victoria}
\affiliation{Nagoya University, Nagoya}
\affiliation{Nara Women's University, Nara}
\affiliation{National Kaohsiung Normal University, Kaohsiung}
\affiliation{National Lien-Ho Institute of Technology, Miao Li}
\affiliation{Department of Physics, National Taiwan University, Taipei}
\affiliation{H. Niewodniczanski Institute of Nuclear Physics, Krakow}
\affiliation{Nihon Dental College, Niigata}
\affiliation{Niigata University, Niigata}
\affiliation{Osaka City University, Osaka}
\affiliation{Osaka University, Osaka}
\affiliation{Panjab University, Chandigarh}
\affiliation{Peking University, Beijing}
\affiliation{Princeton University, Princeton, New Jersey 08545}
\affiliation{University of Science and Technology of China, Hefei}
\affiliation{Seoul National University, Seoul}
\affiliation{Sungkyunkwan University, Suwon}
\affiliation{University of Sydney, Sydney NSW}
\affiliation{Tata Institute of Fundamental Research, Bombay}
\affiliation{Toho University, Funabashi}
\affiliation{Tohoku Gakuin University, Tagajo}
\affiliation{Tohoku University, Sendai}
\affiliation{Department of Physics, University of Tokyo, Tokyo}
\affiliation{Tokyo Institute of Technology, Tokyo}
\affiliation{Tokyo Metropolitan University, Tokyo}
\affiliation{Tokyo University of Agriculture and Technology, Tokyo}
\affiliation{Toyama National College of Maritime Technology, Toyama}
\affiliation{University of Tsukuba, Tsukuba}
\affiliation{Utkal University, Bhubaneswer}
\affiliation{Virginia Polytechnic Institute and State University, Blacksburg, Virginia 24061}
\affiliation{Yokkaichi University, Yokkaichi}
\affiliation{Yonsei University, Seoul}
  \author{J.~Zhang}\affiliation{University of Tsukuba, Tsukuba} 
  \author{M.~Nakao}\affiliation{High Energy Accelerator Research Organization (KEK), Tsukuba} 
  \author{K.~Abe}\affiliation{High Energy Accelerator Research Organization (KEK), Tsukuba} 
  \author{K.~Abe}\affiliation{Tohoku Gakuin University, Tagajo} 
  \author{T.~Abe}\affiliation{Tohoku University, Sendai} 
  \author{I.~Adachi}\affiliation{High Energy Accelerator Research Organization (KEK), Tsukuba} 
  \author{H.~Aihara}\affiliation{Department of Physics, University of Tokyo, Tokyo} 
  \author{M.~Akatsu}\affiliation{Nagoya University, Nagoya} 
  \author{Y.~Asano}\affiliation{University of Tsukuba, Tsukuba} 
  \author{T.~Aso}\affiliation{Toyama National College of Maritime Technology, Toyama} 
% ...
\collaboration{The Belle Collaboration}

\begin{abstract}
We report the first observation of the charmless vector-vector decay
process  $B^+\to\rho^+\rho^0$. The measurement uses a $78\,\rm{fb^{-1}}$
data sample collected with the Belle detector at the KEKB asymmetric
$e^+e^-$ collider operating at the $\Upsilon(4S)$ resonance.
We obtain a branching fraction of ${\cal{B}}
(B^+ \to \rho^+\rho^0)=(31.7\pm 7.1({\rm stat.})^{+3.8}_{-6.7}({\rm
  sys.}))\times 10^{-6}$. 
An analysis of the $\rho$ helicity-angle distributions
gives a longitudinal polarization of $ \Gamma_{L}/\Gamma= 
	(94.8\pm 10.6({\rm stat.})\pm 2.1({\rm sys.}))\%$.
\end{abstract}


\maketitle

Charmless $B$ meson decays to two pseudoscalar mesons (PP) or to
pseudoscalar plus vector meson (PV) final states have been studied in
some detail~\cite{PDG}.  However, measurements of decays to charmless
vector-vector (VV) final states are rather limited; to date only $B
\to \phi K^*$ decays have been observed~\cite{obs-phi-K}. 
The VV decays provide opportunities to search for direct-$CP$ and/or
$T$ violations through angular correlations between the vector meson
decay final states~\cite{br-kramer, a.datta}.
In $B^+\to\rho^+\rho^0$ decays, isospin-breaking processes such as
electroweak penguins~\cite{ewpenguin} or $\rho^0$-$\omega$
interference~\cite{gardner, br-eno}, which may produce a sizable 
direct-$CP$-violating asymmetry, are expected to be enhanced relative 
to $CP$-conserving processes such as gluonic penguins, which are
nominally forbidden by isospin symmetry. The branching fraction for
this process is predicted to be ${\cal O}(10^{-5})$ \cite{br-eno, br-hou}.

In this paper, we present the first observation of the VV decay mode
$B^+\to\rho^+\rho^0$~\cite{charge}.
These decays produce final states where both $\rho$ mesons
are either longitudinally or transversely polarized. We denote the
corresponding amplitudes as $H_{00}$ and $H_{11}$, respectively.  

The analysis is based on a 78~$\rm fb^{-1}$ data sample containing
$85\times10^6$ $B$ meson pairs collected at the $\Upsilon(4S)$
resonance with the Belle detector at the KEKB asymmetric-energy
$e^+e^-$ collider. We also use an off-resonance data sample of
8.3~$\rm fb^{-1}$ collected at a center-of-mass energy that is 60~MeV
below the $\Upsilon(4S)$ resonance.

The Belle detector is a large-solid-angle magnetic spectrometer.
Charged particle tracking is provided by a three-layer silicon vertex
detector and a 50-layer central drift chamber (CDC).
Charged hadron identification is provided by $dE/dx$ measurements in
the CDC, and arrays of aerogel threshold \v{C}erenkov counters (ACC)
and time-of-flight scintillation counters (TOF) that surround the CDC.
An electromagnetic calorimeter comprised of CsI(Tl) crystals (ECL)
provides photon detection and electron identification.  All of these
devices are located inside a superconducting solenoidal coil that
provides a 1.5~T magnetic field. An iron flux-return located outside
of the coil is instrumented to detect $K_L^0$ mesons and muons. 
The detector is described in detail elsewhere
\cite{Belle}.

We select $B^+\to\rho^+\rho^0$ candidate events by combining three
charged pions and one neutral pion.  We require  that each charged
track has a transverse momentum $p_t>0.1{\,{\rm GeV}/c}$, and is
consistent with originating from  within $\delta r<0.1$~cm in the
radial  direction and $|\delta z|<5$~cm in the electron beam direction
of the run-by-run-determined interaction point. We also require that
the three charged tracks be positively identified as pions by the CDC,
ACC and TOF systems.

Candidate $\pi^0$ mesons are reconstructed from pairs of photons with
an invariant mass in the range 
$0.118{\,{\rm GeV}/c^2}<M(\gamma\gamma)<0.150 {\,{\rm GeV}/c^2}$. 
For the ECL barrel region ($32.2^\circ < \theta < 128.7^\circ$),
photon energies greater than $50$~MeV are required; for the ECL endcap
region ($17.0^\circ<\theta<31.4^\circ$ or $130.7^\circ
<\theta<150.0^\circ$), this requirement is increased to $100$~MeV. In
addition, we only accept  $\pi^0$ candidates with a $\Upsilon(4S)$
center-of-mass system (cms) momentum  $p_{\pi^0}>0.5{\,{\rm GeV}/c}$. 
The $\pi^0$ candidates are kinematically constrained to the nominal
$\pi^0$ mass.

Candidate $\rho$ mesons are reconstructed via their 
$\rho^0 \to \pi^+\pi^-$ and $\rho^+ \to \pi^+\pi^0$ decays.
For both the charged and neutral modes, we require 
$0.65{\,{\rm GeV}/c^2}<M(\pi\pi)<0.89{\,{\rm GeV}/c^2}$.

$B^+\to\rho^+\rho^0$ decays are identified using the beam-energy
constrained mass $M_{bc}\equiv \sqrt{(E_{\rm beam})^2-(p_B)^2}$ and
the energy difference $\Delta E\equiv E_B-E_{\rm beam}$, where $E_{\rm
  beam}$ is the cms beam energy, and $p_B$ and $E_B$ are the cms
momentum and energy, respectively, of the $B^+ \to \rho^+\rho^0$
candidates. The $\Delta E$ distribution has a tail on  the lower side
caused by incomplete longitudinal containment of electromagnetic
showers in the CsI crystals, and the $\Delta E$ resolution varies
slightly depending on the $\pi^0$ momentum.  
We select events in the region $|\Delta E|<0.4$~GeV,
$M_{bc}>5.2{\,{\rm GeV}/c^2}$ with a  signal region defined as 
$-0.10{\,\rm GeV}<\Delta E<0.06{\,\rm GeV}$ and 
$5.272{\,{\rm GeV}/c^2}<M_{bc}<5.290{\,{\rm GeV}/c^2}$.  
The $\rho\to\pi\pi$  daughter momentum distributions are
helicity-state dependent: for the $H_{00}$ state, the two pions  have
an asymmetric momentum distribution where one pion has low momentum
(in the range $0 \sim 1.3{\,{\rm GeV}/c}$) while the other has high
momentum ($1.3 \sim 2.8{\,{\rm GeV}/c}$); 
for the  $H_{11}$ state, the two pions tend to have the same
momentum. Because of its higher probability for having a low energy
pion, the $H_{00}$ state has a lower reconstruction efficiency and a
$\Delta E$ resolution that is, on average, about 15\% broader than
that for the $H_{11}$ state.

There are large backgrounds from  $e^+e^-\to q\bar{q}$ continuum
events ($q=u,d,s,c$), which tend to have a two-jet-like structure.
These are suppressed by requiring $|\cos\theta_{\mathrm {thr}}|<0.8$,
where $\theta_{\mathrm {thr}}$ is the angle between the thrust  axis
of the candidate tracks and that of the remaining tracks in the event.
We achieve further suppression by means of a likelihood ratio derived
from a Fisher discriminant formed from six modified Fox-Wolfram
moments~\cite{fox} and $\theta_B$, the angle between the $B$ flight
direction and the electron beam direction.  The combined rejection for
continuum events is 98\%, with a 65\% loss in signal. 

Background contributions from $b \to c$ processes are investigated
with a large sample of Monte Carlo (MC) events, for which no signal-like
peak is found in either the $\Delta E$ or $M_{bc}$ distributions.
Some rare $B$ decay processes, such as $B^+\to \eta'\rho^+$, $K^{*+}
\rho^0$, $\rho ^+ K^{*0}$  and $\rho\pi$, can survive the event
selection but are displaced from the signal in $\Delta E$. 
Moreover, these modes have small branching fractions~\cite{rhopi} and
low reconstruction efficiencies. MC estimates based on measured upper
limits for the branching fractions indicate a possible signal-region
yield from these rare modes of seven events; this is taken into
account in the systematic error determination, as discussed below.

\begin{figure}[htbp]
{\centerline{
\epsfysize 1.7 truein
\epsfbox{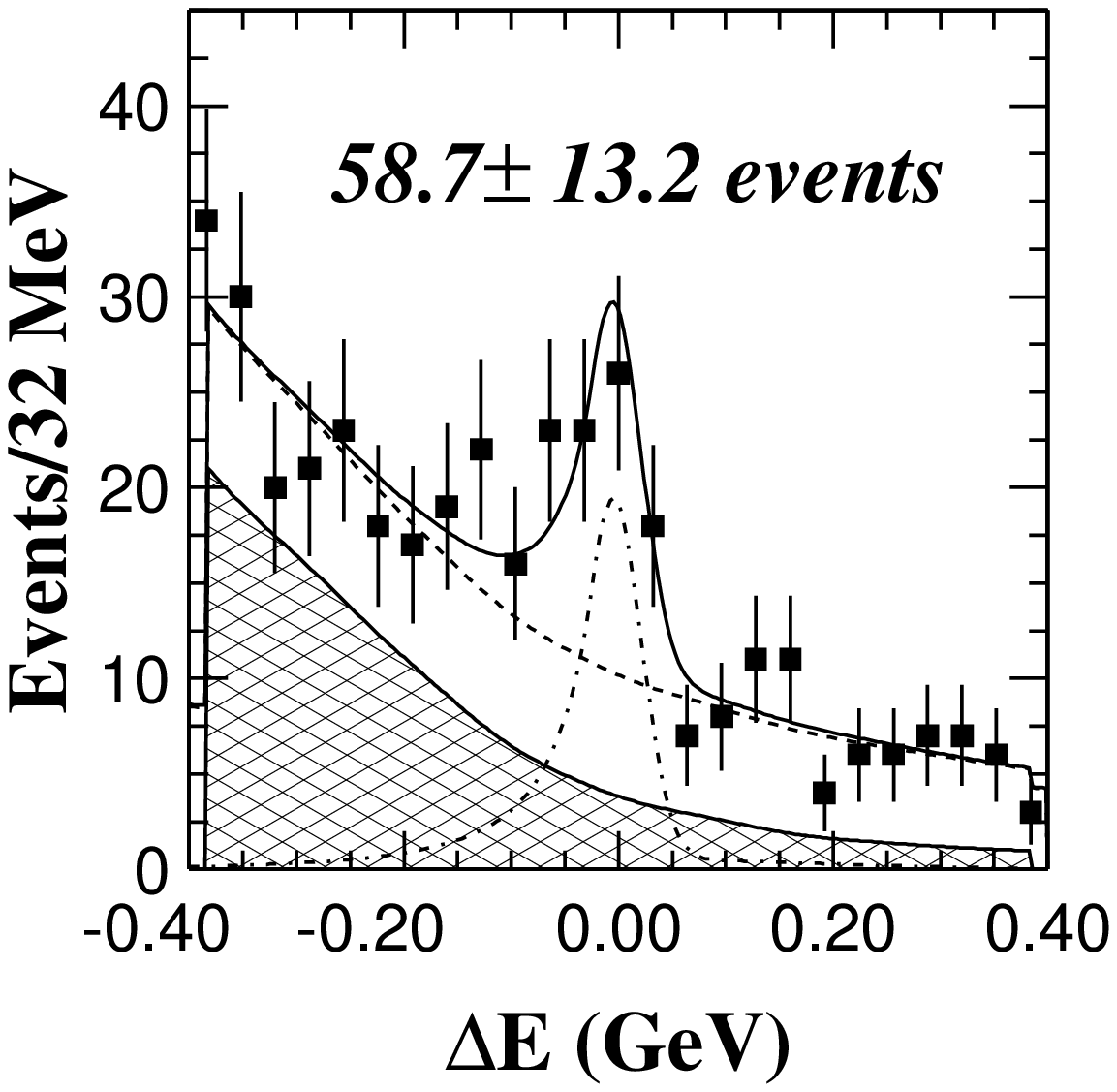}
\epsfysize 1.7 truein
\epsfbox{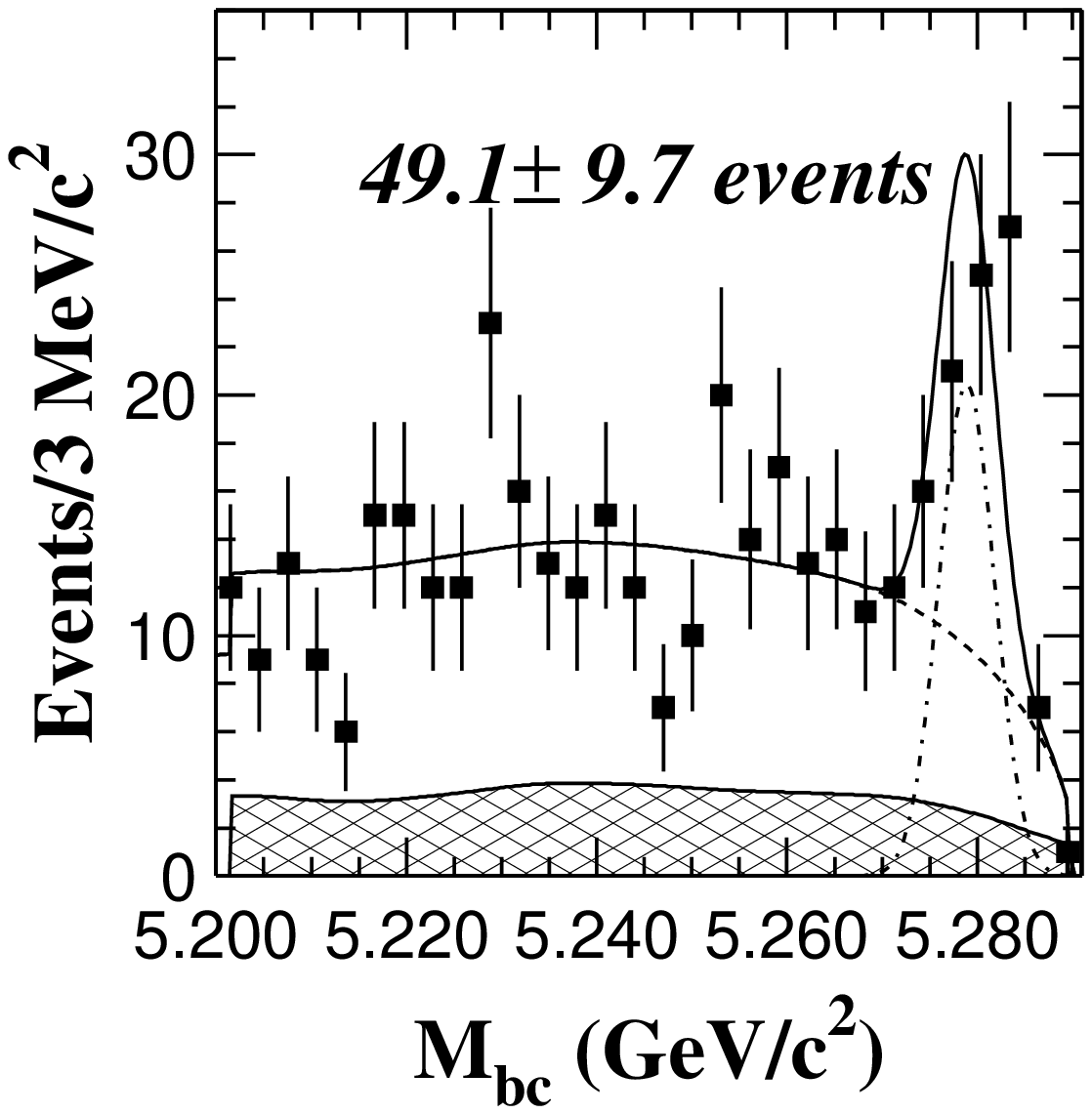}}
\caption{\label{fitdata}
$\Delta E$ (left) and $M_{bc}$ (right) fits to the $B^+\to
  \rho^+\rho^0$ candidate events. The signal component is shown as a
  dot-dashed line. The sum of $B\bar{B}$ and continuum components is
  shown as a dashed line. The shaded histograms represent the
  $B\bar{B}$ background.}}
\end{figure}

Figure~\ref{fitdata} (left) shows the $\Delta E$ projection of
the selected events in the $5.272{\,{\rm GeV}/c^2}<M_{bc}<5.290{\,{\rm
GeV}/c^2}$ signal region.  The curve shows the results of a binned
maximum-likelihood fit with three components: signal, continuum
background, and $B\bar{B}$ background.   The signal is represented by
the sum of a Gaussian and a ``Crystal Ball'' line shape (CB)
function~\cite{cb} with parameters  determined from an
$H_{00}$ signal MC that is calibrated with $B^+\rightarrow\bar{D}^0\pi^+$,
$\bar{D}^0\rightarrow K^+ \pi^- \pi^0$ events.
A linear function with a slope determined from the off-resonance data
is used to represent the continuum background.  The $B\bar{B}$
background contribution is modeled by a smoothed histogram with a
shape that is obtained from MC. In the fit, all parameters other than
the normalizations are fixed. 

The fit gives a signal yield of $58.7\pm13.2 $ events. The statistical
significance of the signal, defined as $\sqrt{-2\ln({\cal L}_0/{\cal
    L}_{\rm max})}$, where ${\cal L}_{\rm max}$ is the likelihood
value at the best-fit signal  yield and ${\cal L}_{0}$ is the value
with the signal yield fixed to zero, is 5.3.

Figure~\ref{fitdata} (right) shows the $M_{bc}$ projection of events
in the $-0.10{\,\rm GeV}<\Delta E<0.06{\,\rm GeV}$ signal region. The
curve shows the results of a binned maximum-likelihood  fit that uses
a single Gaussian with a MC-determined width to represent the signal,
a threshold (ARGUS) function~\cite{argus} for the continuum 
background with shape parameters that are determined from the
$\Delta E$ sideband (defined as $0.1{\,\rm {GeV}}<\Delta E<0.4{\,\rm
{GeV}}$), and a smoothed histogram obtained from MC to represent the 
$B\bar{B}$ background, normalized according to the MC expectation.
This fit gives a signal yield of $49.1\pm 9.7$, with a statistical
significance of 6.5.

\begin{table}[htbp]
\caption {
Signal yields from the fits to the $\Delta E$ and $M_{bc}$
distributions together with the MC-determined efficiencies.}
\label{firstfit}
\begin{tabular}{l c c}\hline \hline
      &        $\Delta E$  fit     &   $M_{bc}$ fit \\ \hline
 Yield  & $58.7\pm13.2$ ($5.3\sigma$)& $49.1\pm9.7$ ($6.5\sigma$)\\
 Efficiency $\epsilon_{00}$    & 2.11\%   & 1.59\% \\
 Efficiency $\epsilon_{11}$    & 3.45\%   & 3.07\% \\ \hline\hline
\end{tabular}
\end{table}

The fit results and MC efficiencies are summarized in
Table~\ref{firstfit}. Since the $\Delta E$ distribution provides
stronger discrimination against rare $B$-meson decay backgrounds, we
use the $\Delta E$ fit result for the branching fraction calculation.

\begin{figure}[htbp]
{\centerline{
\epsfysize 1.6 truein
\epsfbox{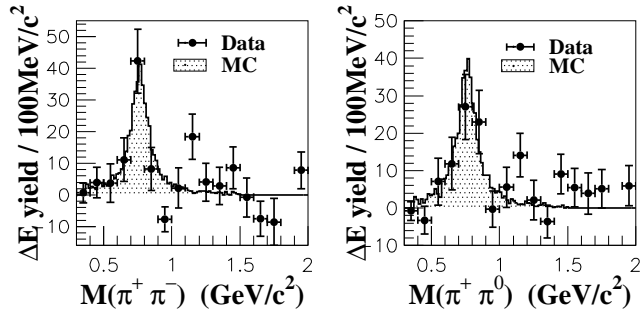}}
\caption {\label{de-rhobin}
Data points are the results of fits to the $\Delta E$ distributions
for each $M(\pi^+\pi^-)$ bin (left) and $M(\pi^+\pi^0)$ bin (right),
where $D\to K\pi$, $\pi\pi$, $\pi\pi^0$ events are excluded by
$-120<M(\pi\pi)-M_{D^0}<30 $ ${\rm MeV}/c^2$  and
$|M(\pi\pi^0)-M_{D^+}|<50$ ${\rm MeV}/c^2$ vetoes; the
histograms are expectations from the signal MC.}}
\end{figure}

Figure~\ref{de-rhobin} shows signal yields extracted from fits to the
$\Delta E$ distributions for different $M(\pi^+\pi^-)$ and
$M(\pi^+\pi^0)$ mass bins; the $\pi\pi$ mass spectra from the signal
MC are shown as shaded histograms. The data agree reasonably well with
$B^+\to\rho^+\rho^0$ MC expectations.

We examined the possible contribution from non-resonant processes
using MC-generated $B^+\to \pi^+\pi^-\pi^+\pi^0$ events where the
final states are distributed uniformly over phase space.  After the
application of all selection requirements, including the
$\rho$ mass cuts, we find an efficiency that is less than $2\%$ of
that for $B^+ \to \rho^+\rho^0$ decays. 
Possible contributions from $B^+\to a_1^+\pi^0$ or $B^+\to
a_1^0\pi^+$ decay are examined and found to be smaller than those from
non-resonant decays. To account for these contributions, we perform
$\chi^2$ fits to the distributions shown in Fig.~\ref{de-rhobin} with
a $\rho$ plus non-resonant $\pi\pi$ component included. The resulting
non-resonant yield increased by $1\sigma$ is included in the
systematic error.

\begin{figure}[htbp]
{\centerline{
\epsfysize 1.7 truein
\epsfbox{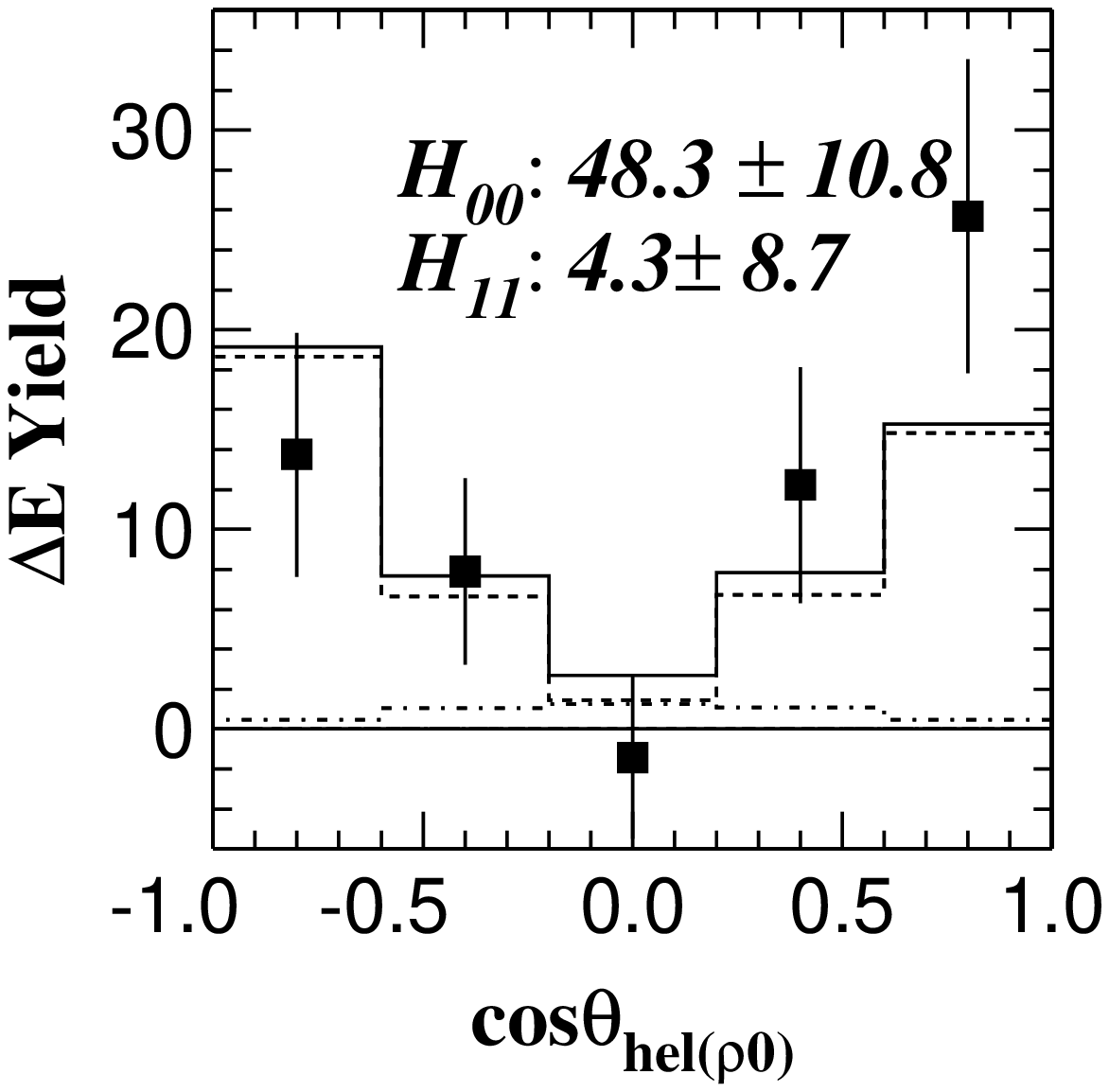} 
\epsfysize 1.7 truein
\epsfbox{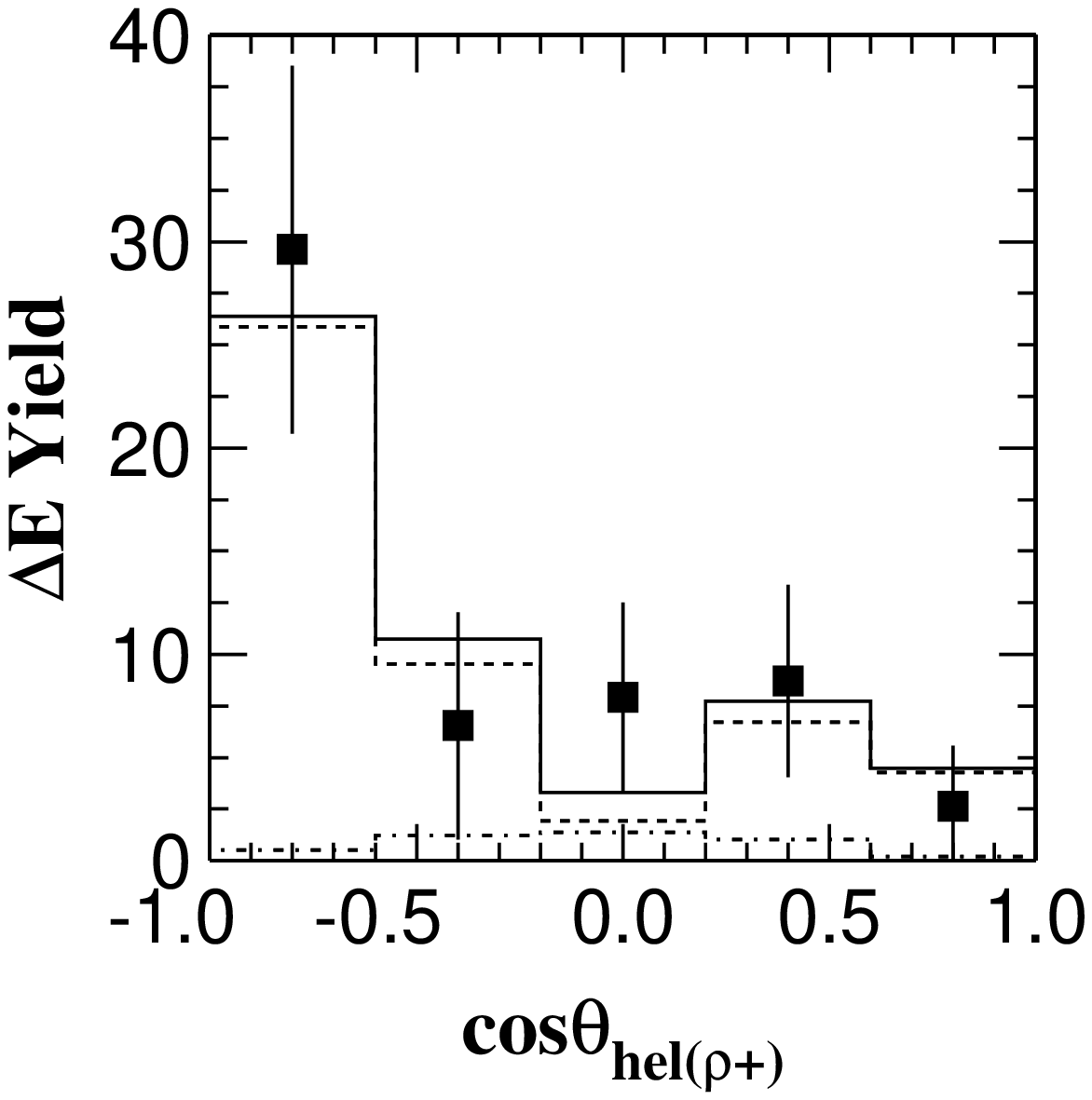}}
\caption{\label{simu-fit}
Data points show the background-subtracted $\cos\theta_{\rm hel}$
distributions for the $\rho^0$  (left) and $\rho^+$ (right). In each
plot the dashed (dot-dashed) histogram is the $H_{00}$ ($H_{11}$)
component of the fit; the solid histogram is their sum. 
The low yield of events near $\cos\theta_{{\rm hel}(\rho^+)}=1$ is due
to the $p_{\pi^0}>0.5{\,\rm GeV}/c$ requirement.}}
\end{figure}

We use the $\rho\to\pi\pi$ helicity-angle ($\theta_{\rm {hel}}$)
distributions to determine the relative strengths of $H_{00}$ and
$H_{11}$.  Here $\theta_{\rm {hel}}$ is the angle between the $\rho$
flight direction in the $B$ rest frame and $\pi^+$ flight direction in
the $\rho$ rest frame. The signal yields determined from fits to the
$\Delta E$ distributions for each helicity-angle bin are plotted
versus $\cos\theta_{\rm hel}$ in Fig.~\ref{simu-fit} for the $\rho^0$
(left) and the $\rho^+$ (right). In determining these yields,
bin-by-bin $H_{00}$-signal-MC-determined shape parameters are used to
extract the signal. We perform a simultaneous $\chi^2$ fit to the two
background-subtracted $\rho$ helicity-angle distributions using
MC-determined  expectations for the $H_{00}$ and $H_{11}$ helicity
states. The fit results, shown as histograms in Fig.~\ref{simu-fit},
indicate that the longitudinal ($H_{00}$) state dominates. We obtain
the acceptance-corrected longitudinal polarization ratio
	$$ \frac{\Gamma_{L}}{\Gamma}= 
	(94.8\pm 10.6({\rm stat.})\pm 2.1({\rm sys.}))\%, $$
where the systematic error includes uncertainties in the
signal yield extraction and the polarization-dependence of
the detection efficiency. This dominance of $H_{00}$ is consistent 
with theoretical predictions~\cite{longitudinal}.

We use MC-determined efficiencies for the two helicity states and the
measured polarization ratio to determine the branching fraction.
We assign a 3.4\% systematic error for tracking that is obtained from
a study of partially reconstructed $D^*$ decays; 
a 3.6\% error for the particle identification efficiency that is based
on a study of kinematically selected $D^{*+}\to D^0\pi^+$, $D^0\to
K^-\pi^+$ decays; 
a 4.0\%  $\pi^0$ reconstruction systematic error that is determined
from comparisons of $\eta\to \pi^0\pi^0\pi^0$ to $\eta
\to\pi^+\pi^-\pi^0$ and  $\eta \to \gamma\gamma$; 
a  5.4\% error for continuum suppression that is estimated from a
study of $B^+\rightarrow \bar{D}^0\pi^+$, $\bar{D}^0\rightarrow K^+
\pi^- \pi^0$ decays; 
an error associated with the $\Delta E$ fit of $^{+7.3}_{-6.7}\%$ that
is obtained from changes that occur when each parameter of the fitting
functions is varied by $\pm 1 \sigma$;
a 1\% error for the uncertainty in the number of $B\bar{B}$ events in
the data sample; 
a $^{+0}_{-16.7}\%$ error to account for a possible contribution from
non-resonant decays;
and a 3.3\% error due to uncertainties in the rare $B$ decay
background that is estimated from the change produced by fitting the
$\Delta E$ distribution with the inclusion of an additional component
normalized at its MC expectation.
We also include a $^{+3.1}_{-6.5}\%$ error due to the uncertainty in
the mixture of helicity states that is obtained from the changes in
the branching fraction that occur when the longitudinal polarization
ratio is lowered by $1\sigma$ or raised to  $100$\%.
The quadratic sum of all of these errors is taken as the total
systematic error.  We obtain the branching fraction
$${\cal B}(B^+\to \rho^+
\rho^0)=(31.7\pm 7.1({\rm stat.})^{+3.8}_{-6.7}({\rm
sys.}))\times 10^{-6}.$$ 

As a check, we examined the decay mode $B^+ \to \rho^+ \bar {D^0}$,
$\bar{D^0} \to \pi^+\pi^-$, which has the same final state particles as
the mode under study, including a $\pi^0$ with a similar momentum
distribution.  The same analysis procedure is applied  except for an
$|M(\pi\pi)-M_{D^0}|<13$~${\rm MeV}/c^2$ mass selection.  For this mode,
we obtain a signal yield of $42.3\pm8.5$ events, consistent with MC
expectations based on the known branching fraction values~\cite{PDG}.

Direct $CP$ violation would be indicated by a difference in partial
rates for $B^-\to\rho^-\rho^0$ and $B^+\to\rho^+\rho^0$. Separate fits
to the $\Delta E$ distributions find $29.3\pm 9.5$ $\rho^- \rho^0$
events and $29.3\pm 9.1$ $\rho^+\rho^0$ events. Since backgrounds from
generic $B\bar B$ decays should contribute equally to $\rho^-\rho^0$
and $\rho^+\rho^0$, we fix the normalizations for $B\bar B$ components
at half the value determined from the combined fit. 

The charge symmetry of the detector and reconstruction procedure is
verified with a sample of $B^{+}\to \bar{D}^0\pi^{+}$, $\bar{D}^0\to
K^+\pi^-\pi^0$  decays and their charge conjugates. 
Here the analysis procedure is similar to that for $B^+\to \rho^+\rho^0$ 
but replacing one $\pi^+$ by a $K^+$ and the invariant mass
requirements by $|M(K\pi\pi^0) - M_{D^0}|<50\,{\rm{MeV}}/c^2$.
For these events we find a partial rate asymmetry of $(-2.1\pm2.5)\%$,
which is consistent with zero. We assign $2.5\%$ as the systematic
error for the detection and reconstruction  asymmetry.
The systematic error associated with the $\Delta E$
fitting procedure is determined to be ($^{+0.8}_{-1.2}$)\% 
by shifting each parameter of the fitting functions by $\pm
1\sigma$ and taking the quadratic sum of the resulting changes in
${\cal A}_{CP}$. The quadratic sum of these errors is taken as the
total systematic error. We obtain the ${CP}$ asymmetry 
\begin{align*}
{\cal A}_{CP}( B^{\mp}\to \rho^{\mp}\rho^0 ) &
       \equiv \frac{N_{(\rho^-\rho^0)}-N_{(\rho^+\rho^0)}}
        {N_{(\rho^-\rho^0)}+N_{(\rho^+\rho^0)}} \\
        & =0.00 \pm 0.22({\rm stat.}){\pm 0.03}({\rm sys.}).
\end{align*}

In summary, we have observed the decay $B^{+}\rightarrow \rho^+
\rho^{0}$ with a statistical significance of 5.3. We measure the
branching fraction to be ${\cal{B}} (B^+\to \rho^+\rho^0)=(31.7\pm
7.1({\rm stat.})^{+3.8}_{-6.7}({\rm sys.}))\times10^{-6}$, where the
systematic error includes the error associated with the  helicity-mix
uncertainty. An analysis of the helicity-angle distributions gives the
longitudinal polarization ratio 
$ \Gamma_{L}/{\Gamma}= 
        (94.8\pm 10.6({\rm stat.})\pm 2.1({\rm sys.}))\%$.
We also measure
${\cal A}_{CP}( B^{\mp}\to \rho^{\mp}\rho^0 )= 0.00 \pm 0.22({\rm
  stat.}){\pm 0.03}({\rm sys.})$.\\

\begin{acknowledgments}
We wish to thank the KEKB accelerator group for the excellent
operation of the KEKB accelerator.
We acknowledge support from the Ministry of Education,
Culture, Sports, Science, and Technology of Japan
and the Japan Society for the Promotion of Science;
the Australian Research Council
and the Australian Department of Industry, Science and Resources;
National Science Foundation of China under contract No.~10175071;
the Department of Science and Technology of India;
the BK21 program of the Ministry of Education of Korea
and the CHEP SRC program of the Korea Science and Engineering
Foundation; the Polish State Committee for Scientific Research
under contract No.~2P03B 01324;
the Ministry of Science and Technology of Russian Federation;
the Ministry of Education, Science and Sport of Slovenia;
the National Science Council and the Ministry of Education of Taiwan;
and the U.S. Department of Energy.
\end{acknowledgments}


\begin{thebibliography}{99}
\bibitem{charge} The inclusion of charge conjugate modes is implied
unless stated otherwise.
\end{thebibliography}
\end{document}